# Optimizing the Total Production and Maintenance Cost of an Integrated Multi-Product Process and Maintenance Planning (IPPMP) Model

Mohammad Arani
Department of Systems Engineering
The University of Arkansas at Little Rock
Little Rock, the USA
mxarani@ualr.edu

Mousaalreza Dastmard
Department of Industrial and Civil Enginering
The Sapienza University of Rome
Lazio, Italy
dastmard.1852433@studenti.uniroma1.it

Zhila Dehdari Ebrahimi
Department of Transportations and Logistics
The North Dakota State University
Fargo, the USA
zhila.dehdari@ndsu.edu

Mohsen Momenitabar
Department of Transportations and Logistics
The North Dakota State University
Fargo, the USA
mohsen.momenitabar@ndsu.edu

Xian Liu
Department of Systems Engineering
The University of Arkansas at Little Rock
Little Rock, the USA
xxliu@ualr.edu

*Abstract*—Today, a competitive manufacturing environment imposes further production cost reduction on modern companies. Seeking proper recommendations in production and maintenance planning are the two essential cornerstones of effective production organizations. In the current research, we have considered the problem of integrated multi-product process and maintenance planning on a capacitated machine that is susceptible to random breakdown. Maintenance processes comprise general perfect repair (non-cyclical) as preventive maintenance (PM) in the early stages and minimal repair as corrective maintenance for the occurrence of machine breakdown. Furthermore, a rational presumption is reflected in the problem statement in which the time and cost of PM are pertinent to the interval between the prior perfect repair and current PM. The purpose served by this paper is to minimize the cost of production accompanying PM, and the expected corrective repair, consequently, a mixed-integer linear programming (MILP) model has been constructed to pave the way. The model investigated under two circumstances of the machine age effect and its absence. The outcome depicted that the presence of the machine age effect led to an accurate and lessen total cost calculation.

## I. Introduction

Manufacturing industries continually deal with two imperative subjects: production, and maintenance planning. The Production department seeks to minimize production costs typically including production, holding, backorder, and setup cost. Production managers look for maximum machinery capacity, and key equipment to meet customers' demands within a promised level of quality [1]. On the other hand, the maintenance department has to perform preventive maintenance and adopts maintenance strategies like dynamic sampling strategy [2], semi-dynamic maintenance [3], dynamic maintenance [4], and pro-active maintenance scheduling [5] to keep machines functionality well and prevent the failure of the machine due to breakdown.

Both production and maintenance departments' activities have been done on the same equipment and must use equipment capacity to promote productivity and reliability of them. Although carrying out PM may prevent failure, because of the differences in the objectives of these two departments, the conflict arises [6]. One would like to produce non-stop and the other is likely to put more weight on longevity, reliability, and required service level of the equipment [7]. Notwithstanding the trade-off between the activities of production scheduling and maintenance planning, they are classically planned and executed individually in a manufacturing system even if industrial productivity can be improved by optimization of both production scheduling and maintenance planning decisions concurrently [8].

Therefore, it is necessary to find an optimal balance of production scheduling and maintenance activities for the equipment. So, for industrial factories, coordination between production and maintenance departments is essential to prevent production interruption and unplanned repair (corrective maintenance).

This paper is organized as follows: in the next section, we considered the literature review dividing into three groups, such as lot size, maintenance, integrated production, and maintenance planning. In section 3, we specify the innovation that makes the paper distinguished. Then the problem statement and mathematical formulation are described in sections 4 and 5 respectively. The methodology and software used to solve the problem are mentioned in section 6. Section 7 is dedicated to numerical examples and the results. Finally, the conclusion and potential future work are presented in section 8.

## II. LITERATURE REVIEW

In this section, we will review the past work which has been done by scholars. In this way, two categories have been considered including (A) lot size and maintenance, and (B) Integrated production and maintenance planning.

### A. Lot Size and Maintenance

Most researchers in the world are working on lot sizing. The majority of research is about the multi-level capacitated lot-sizing problem such as [9] and [10]. All economic lot-sizing problems are seeking for minimizing costs while satisfying demand. Andriolo et al [11] did comprehensive surveys from 1913 to 2014. Lot size has a direct effect on scheduling plans. For this reason, [11] and [12] considered the problem of production planning and scheduling simultaneously. Faccio et al [13] covered a variety of maintenance strategies to reduce spare parts cost, human resources, missing production capacity, and other indirect costs. Kader et al [14] studied the problem of spare parts in maintenance planning and considered the use and new spare parts to be replaced during corrective and preventive maintenance.

### B. Integrated production and maintenance planning

Budai et al [15] surveyed the papers considering the relation between production and maintenance planning. In this paper, three relationships were mentioned: Production planning which is maintenance based, maintenance planning which is production based and integrated production and maintenance planning. Aghezzaf and Najid [16] proposed two models for the parallel manufacturing system. On failure, maintenance policy consists of minimal repair to restore the machine to the previous working condition. The perfect repair is performed as preventive maintenance approach to return the machine to as good as a new one. The model of periodic PM is mixed-integer nonlinear programming and the model of non-periodic PM planning is mixed-integer linear programming. A Lagrange-based heuristic method was used to solve the problems. Yalaoui et al [17] extended the solution method by using the polyhedral theory which is developed to reduce the time of solving a wide range of large problems in an acceptable period. Aghezzaf et al [18] presented an integrated production and maintenance planning model considering imperfect preventive maintenance. The imperfect repair reduces the machine age and does not restore it to a new one condition. Mixed-Integer linear programming was proposed for this model. Nourelfath and Chatelet [19] considered a manufacturing system with parallel components in which multi-component PM has less cost than single component PM. The aim is to reduce the sum of PM cost, corrective repair cost, production cost, holding, shortage, and setup cost. Fakher et al [20] used a hybrid genetic algorithm to solve the problem and hybrid GA with the Tabu Search algorithm to manage the initial GA population. Nourelfath et al [21] involved quality in production and maintenance planning. The problem is considered as multi-product and multi-period lot-sizing in which machines have two states of under control and out of control. In out of control state, some of the products are not acceptable. The purpose is to minimize the total cost while satisfying customers' demand. Results show that if PM cost is not too high, carrying out PM can reduce the quality cost. Also, general (not necessarily periodic) PM reduces total costs. Likewise, Bouslah et al [22] considered interactions between production, holding, maintenance, and quality by continuous sampling. Bouslah et al [23] also considered three aspects of integrated production planning, maintenance planning, and quality control. Dahane et al [24] focused on the relation between machine failure and production rate. They considered two product types. Type a demand must be satisfied without any shortage over the planning horizon and satisfying demand of type b product is optional though its profit is high. The aim is to maximize the total expected profit. Chelbi et al [25] combined lot-sizing and maintenance problem regarding the defective items. The goal is to find optimal PM length and reducing the cost of finished goods.

In general, the noncyclical PM has better performance than periodic one. Gustavsson et al [26] proposed a binary linear programming model for multi-component maintenance planning regarding the dependency of PM cost on the interval. This paper proved that considering dependency improves maintenance costs.

## III. INNOVATION

Aghezzaf and Najid [16] modeled production and maintenance planning with noncyclical PM. Although the model is mixed-integer linear programming and includes setup cost, this model has not considered the cost of corrective repair to zero if the machine does not run and machine age remains constant. The same drawback is seen in [17] and [6]. In practice, a significant percentage of machine capacity is allocated to set up activities. Industries need a more realistic and comprehensive model to plan [27]. In this study, in addition to covering the issues stated above, the time and cost of PM are considered based on the interval between the previous PM and the current PM. By increasing this length, the time and cost of PM rise up [26]. The following model has developed the model with the assumption in [26] that was proposed in the form of mixed-integer linear programming. It enables us to reach a globally optimal solution even for large scale problems. The model is non-cyclic and is flexible enough to become periodic by adding a few simple mathematical equations.

## IV. PROBLEM STATEMENT

Assume a capacitated machine in a manufacturing system that faces random failure [28]. The scheduling department including the production and maintenance department seeks plans for optimal production and Maintenance. The maintenance strategy considers corrective maintenance (unplanned repair) which is minimal and restores the machine's previous working condition and does not change its age. If PM is not performed and the machine is used for the production process in a period, the machine age will increase for one period (time unit is period) and failure rate goes up [4], [29]. Maintenance policies focus on noncyclical PM, though it can be periodic. If a perfect PM (replacement maintenance) is conducted, the machine age status is transformed into a new machine meaning the machine age is reset to zero. It is similar to the substitution of the machine with a new one [30].

The cost and time of PM depend on the machine age [31]. Obviously, as machine age increases, the costs and time needed to perform PM increases, because more spare parts are required. Maintenance costs include the overall cost of the PM and planned maintenance costs. The maintenance department's objective is to minimize total maintenance costs.

The production department must satisfy the demand of P product(s) over planning horizon H. To meet the demand, the production department uses the machine capacity to advance the production progress. If in a time period shortage occurs, it is unacceptable, but the demand can be satisfied after the due date with penalty cost. Machines must be prepared for manufacturing products, so setup cost and time must be considered. The production department aims to meet all demand over planning horizon H while minimizing production cost, holding cost, backorder, and setup cost. The scheduling department's mission is combining production and maintenance purposes in the form of a single goal to decrease the overall cost. Each department uses the machine capacity to advance the production and maintenance activities concerning its limitations. The scheduling department should integrate planning to minimize the total production and maintenance costs.

## V. MATHEMATICAL FORMULATION

In this section, we will build our model according to the expected number of failure E, the function of machine failure rate (r), expected cost (ERC), and time (ERT) of the corrective maintenance which we calculated these items separately before using in our model. The maintenance department deals with two types of activities including preventive maintenance that must be done at the start of a period and corrective maintenance that must be done during a period when facing machine failure. Due to the random nature of corrective maintenance, we will use the expected time and cost of corrective maintenance. First, we assume the planning horizon (H) consists of T periods with length ($\tau$). Failure rate remains unchanged with performing the corrective maintenance because it is a minimal repair. So, the expected number of failure E during $[a, b]$ using a non-homogeneous Poisson process is calculated in the below formula:

$$E = \int_a^b r(u)du \quad (1)$$

In which, r is the machine failure rate function obtaining as an equation from density f and cumulative probability function F (equation 2).

$$r(u) = \frac{f(u)}{1 - F(u)} \quad (2)$$

We use equation 1 to calculate the expected cost (ERC) and time (ERT) of the corrective maintenance.

$$ERC = E \times RC \quad (3)$$

$$ERT = E \times RT \quad (4)$$

RC and RT are mean of corrective maintenance cost and time respectively. By setting a and b to machine age at the beginning and end of periods, we can calculate the expected cost and time of corrective maintenance based on the machine age.

### A. Model Formulation

In this section, we will present the model including indices, parameters, and variables which widely used in our model. Based on the, we have:

*Indices:*

$t$: Period index $t \in \{1,2,...,T\}$

$l$: Length index (a period) $l \in \{1,2,...,T\}$

$p$: Product index $p \in \{1,2,...,P\}$

*Parameters:*

$d_{pt}$: The demand for product $p$ in period $t$

$h_p$: A period holding cost of product $p$

$b_p$: A period backorder cost of product $p$

$sc_p$: Machine setup cost for product $p$

$st_p$: Machine setup time for product $p$

$\pi_p$: Process cost of product $p$

$\phi_p$: Required capacity to process product $p$

$PMC^l$: Preventive maintenance costs when the machine age is $l$ at the beginning if the period

$PMT^l$: Preventive maintenance time when the machine age is $l$ at the beginning if the period

$RC$: Corrective maintenance cost

$RT$: Corrective maintenance time

$e^l$: Expected number of machine failures during the period when the machine age is $l$ at the beginning of the period.

$L$: Nominal machine capacity

$M$: A big enough positive number

*Variables:*

$a_t$: (Integer) Machine age at the beginning of period $t$

$z_t^l$: (Binary) If PM set to be done in period $t$ while the previous PM has been done 1 period before $t$, the value is 1 otherwise 0

$Z_t$: (Binary) If PM set to be done in period $t$, the value is 1 otherwise 0

$CPM_t$: (Continuous) PM cost in period $t$

$CRM_t$: (Continuous) Expected corrective maintenance cost in period $t$

$TPM_t$: (Continuous) PM time in period $t$

$TRM_t$: (Continuous) Expected corrective maintenance time in period $t$

$x_{pt}$: (Integer) Production amount of product $p$ in period $t$

$I_{pt}$: (Integer) Inventory level of product $p$ in period $t$

$B_{pt}$: (Integer) Backorder level of product $p$ in period $t$

$y_{pt}$: (Binary) If the machine runs to produce product $p$ in period $t$, the value is 1 otherwise 0

$Y_t$: (Binary) If the machine runs to produce in period $t$, the value is 1 otherwise 0

$w_t^l$: (Binary) If $l = a_t + 1$, then the value is 1 otherwise 0

$E_t$: (Continuous) Expected number of machine failure in period $t$

### B. Problem formulation

In this section, we will build our model based on the variables and parameters which we have defined. The objective function of our model is constituted by two section which is related to the manufacturing cost including the production cost, holding cost, backorder cost, and setup cost. The second term of the objective function is maintenance cost including the PM cost and expected corrective cost which we have calculated on the section 5. The presented model as follow:

$$\text{Min} \sum_{p=1}^{P}\sum_{t=1}^{T}\left(h_p I_{pt} + b_p B_p t + \pi_p x_{pt} + sc_p y_{pt}\right) + \sum_{t=1}^{T}\left(CPM_t + CRM_t\right) \quad (5)$$

$$I_{pt} - B_{pt} = I_{p(t-1)} - B_{p(t-1)} + x_{pt} - d_{pt}, \quad p = 1,...,P, t = 1,...,T \quad (6)$$

$$x_{pt} \leq M y_{pt}, \quad p = 1,...,P, t = 1,...,T \quad (7)$$

$$\sum_{p=1}^{P} \Phi_p x_{pt} \leq L - TPM_t - TRM_t - \sum_{p=1}^{P} st_p y_{pt}, \quad t = 1,...,T \quad (8)$$

$$\sum_{l=1}^{T} z_t^l \leq 1, \quad t = 1,...,T \quad (9)$$

$$Z_t = \sum_{l=1}^{T} z_t^l, \quad t = 1,...,T \quad (10)$$

$$\sum_{t'=1}^{t}\sum_{l=1}^{T} l z_{t'}^l = t - 1, \text{ if } Z_t = 1, \ t = 2,...,T \quad (11)$$

$$a_t = \left(a_{t-1} + Y_t\right)\left(1 - Z_t\right), \quad t = 1,...,T \quad (12)$$

$$MY_t \geq \sum_{p=1}^{P} y_{pt}, \quad t = 1,...,T \quad (13)$$

$$\sum_{l=1}^{T} l w_t^l = a_t + 1, \quad t = 1,...,T \quad (14)$$

$$\sum_{l=1}^{T} w_t^l = 1, \quad t = 1,...,T \quad (15)$$

$$E_t = \left(\sum_{l=1}^{T} e^l w_t^l\right) Y_t, \quad t = 1,...,T \quad (16)$$

$$CPM_t = \sum_{l=1}^{T} PMC^l z_t^l, \quad t = 1,...,T \quad (17)$$

$$CRM_t = E_t RC, \quad t = 1,...,T \quad (18)$$

$$TPM_t = \sum_{l=1}^{T} PMT^l z_t^l, \quad t = 1,...,T \quad (19)$$

$$TRM_t = E_t RT, \quad t = 1,...,T \quad (20)$$

Equation 5 is the objective function with two terms that should be minimized. The first term relates to manufacturing cost including the production cost, holding cost, backorder cost, and setup cost. The second term relates to the maintenance cost including the PM cost and expected corrective cost. Equation 6 balance between the amount of production, demand, and inventory and backorder level. Constraint 7 ensures that the production is enabled while the machine is set up to produce. The limitation of Capacity has been defined as equation 8. Available production capacity is achieved after deducting the capacity requiring for PM activities, downtime corrective maintenance, and setup time from nominal capacity. Equation 6-8 is related to the production part. Equation 9 ensures that not more than one PM has been considered for the machine. Equation 10 defines in which periods PMs are planned to do. Equation 11 ensures that the sum of the intervals between PMs before planned PM in period t must be $t - 1$ periods. The machine age calculation is based on machine run variable $Y_t$ by equation 12. In this equation, the machine age increases if $Y_t = 1$. Equation 13 force $Y_t = 1$ if the machine runs to produce at least one product. Equation 14-16 have allocated the expected number of machine failure in period t based on the machine age. Equation 17 and 19 calculate the cost and time of PM for period $t$. Equation 18 and 20 calculate the expected cost and time of corrective maintenance for period $t$.

### C. Linearization

Equation 11, 12, and 16 are nonlinear. So, they should be converted to linear by replacing linear substitution equations. Equation 11 is determined based on $Z_t$ value. The model has this constraint when $Z_t = 1$. Replacing the following equations makes it linear.

$$\sum_{t'=1}^{t}\sum_{l=1}^{T} l z_{t'}^l \leq t - 1 + \left(1 - Z_t\right) M, \quad t = 1,...,T \quad (21)$$

$$\sum_{t'=1}^{t}\sum_{l=1}^{T} lz_{t'}^{l} \geq t-1-(1-Z_t)M, \ t=1,...,T \quad (22)$$

Equation 12 is the product of two terms. The term $1-Z_t$ is binary, so the equation can be replaced with three following linear equations.

$$a_t \leq (1-Z_t)M, \ t=1,...,T \quad (23)$$

$$a_t \leq (a_{t-1}+Y_t), \ t=1,...,T \quad (24)$$

$$a_t \geq (a_{t-1}+Y_t)-Z_tM, \ t=1,...,T \quad (25)$$

Equation 16 is also the product of a term in a binary variable ($Y_t$). Similarly, it can be replaced by the following linear equations.

$$E_t \leq \left(\sum_{l=1}^{T} e^l w_t^l\right)+(1-Y_t)M, \ t=1,...,T \quad (26)$$

$$E_t \geq \left(\sum_{l=1}^{T} e^l w_t^l\right)-(1-Y_t)M, \ t=1,...,T \quad (27)$$

$$E_t \leq Y_tM, \ t=1,...,T \quad (28)$$

By substitution of equations 21-28 instead of 11, 12, and 16, the nonlinear model is transformed into a linear form.

### D. Periodic Maintenance

The mathematical programming of the model is flexible enough that it can be periodic by adding a few constraints. We define a binary variable $PL^l$ that is 1 if the interval between consecutive PMs is l. The constraints are formulated as follow:

$$\sum_{t=2}^{T} z_t^1 = PL^1(T-1) \quad (29)$$

$$\sum_{t=1}^{T} z_t^l = PL^l\left\lceil\frac{T}{l}-1\right\rceil, \ l=2,...,T \quad (30)$$

$$\sum_{l=1}^{T} PL^l = 1 \quad (31)$$

Equation 29 and 30 ensure the possibility of PMs with interval 1. Equation 29 is somewhat different from 30 because of the assumption that in all circumstances in the first period, PM is carried out and we suppose that the machine is as a new one. Equation 31 ensures that the interval between preventive maintenances is unique. By amending equation 29-31, the periodic assumption would be added to the model.

## VI. METHODOLOGY

It has been tried to obtain a linear model, albeit with integer and binary variables. The advantage of a linear model in comparison to other similar nonlinear models is the certainty to achieve an optimal solution through branch and bound (B&B) procedure by software such as CPLEX and GAMS. But solving nonlinear models and ensuring their global optimality is difficult to achieve. However, non-linear models allow the modeling process to add easier and more realistic assumptions. In this study, the model will be transformed into the linear model so that we can reach to the optimum value easily. Then for solving the model, GAMS software has been used on a laptop with a dual-core 2.5 MHz processor and 4GB of RAM. Finally, the result of the GAMS software has brought on the section 7.

One of the advantages of using the GAMS software is the shorter process time of solving the model. Other advantages are that this software can help us to reach to the optimum value of variables which being defined in our model. When our built model converted to the linear model, this give us the approach for solving the proposed model easily.

## VII. NUMERICAL EXAMPLE

Most of the parameters which we assumed to build our model were taken from [1]. Consider household appliances production factory with a machine that is prone to random failure over an 8-month horizon. The machine is brand new and unused at the beginning of the horizon. This condition is like carrying out PM for the first period. The machine probability function is Weibull (2, 2) and failure rate, expected cost, and time of PM are three parameters that are obtained from the probability function. Table 1 shows the expected numbers of failure, cost, and time of PM based on the machine age. Table 2 contains information about the nominal machine capacity and cost and time of corrective maintenance. The demand is assumed certain and it is shown in Table 3 for two products. The production cost, production time, holding cost, backorder cost, and setup cost and time are arranged in Table 4.

TABLE I. FAILURE RATE, COST OF PM, TIME OF PM

| l | 1 | 2 | 3 | 4 | 5 | 6 | 7 | 8 |
|---|---|---|---|---|---|---|---|---|
| $e^l$ | 0.25 | 0.75 | 1.25 | 1.75 | 2.25 | 2.75 | 3.25 | 3.75 |
| $PMC^l$($) | 1613 | 2016 | 2520 | 3150 | 3937 | 4922 | 6152 | 7690 |
| $PMT^l$(h) | 1.6 | 2.0 | 2.5 | 3.2 | 3.9 | 4.9 | 6.2 | 7.7 |

TABLE II. NOMINAL CAPACITY, CORRECTIVE MAINTENANCE COST, CORRECTIVE MAINTENANCE TIME

| L (h) | RC ($) | RT (h) |
|---|---|---|
| 200 | 2000 | 12 |

TABLE III. DEMAND FOR THE PRODUCT FOR OVER 8 MONTHS

| T | | 1 | 2 | 3 | 4 | 5 | 6 | 7 | 8 |
|---|---|---|---|---|---|---|---|---|---|
| Product | 1 | 22 | 22 | 22 | 22 | 23 | 22 | 20 | 20 |
| | 2 | 25 | 25 | 22 | 25 | 23 | 22 | 20 | 20 |

Assume two models, A and B. Model A (dependency) considers that the time and cost of PM are pertinent to the machine age. In model B (independence), the cost and time of PM are considered independent and are set on average.

TABLE IV. PARAMETERS RELATED TO PRODUCTION PART (HOLDING COST, BACKORDER COST, SETUP COST, SETUP TIME, PRODUCTION COST, AND PRODUCTION TIME)

| Product | $h_p$ | $b_p$ | $sc_p$ | $st_p$ | $\pi_p$ | $\phi_p$ |
|---|---|---|---|---|---|---|
| 1 | 40 | 240 | 1000 | 10 | 90 | 3.6 |
| 2 | 40 | 240 | 1000 | 10 | 90 | 3.6 |

TABLE V. COST AND TIME OF PM IN DEPENDENCY AND INDEPENDENCY

| $l$ | 1 | 2 | 3 | 4 | 5 | 6 | 7 | 8 | Avg. |
|---|---|---|---|---|---|---|---|---|---|
| Dependency (Model A) | | | | | | | | | |
| $PMC^l(\$)$ | 1613 | 2016 | 2520 | 3150 | 3937 | 4922 | 6152 | 7690 | 4000 |
| $PMT^l(h)$ | 1.6 | 2.0 | 2.5 | 3.2 | 3.9 | 4.9 | 6.2 | 7.7 | 4 |
| Independency (Model B) | | | | | | | | | |
| $PMC^l(\$)$ | 4000 | 4000 | 4000 | 4000 | 4000 | 4000 | 4000 | 4000 | 4000 |
| $PMT^l(h)$ | 4 | 4 | 4 | 4 | 4 | 4 | 4 | 4 | 4 |

TABLE VI. OPTIMAL SOLUTION RESULTS FOR MODEL A AND B (PART A)

| | $CPM_t$ | | | | | | | |
|---|---|---|---|---|---|---|---|---|
| Period (t) | 1 | 2 | 3 | 4 | 5 | 6 | 7 | 8 |
| Model A | 0 | 1613 | 0 | 2016 | 0 | 2016 | 0 | 0 |
| Model B | 0 | 0 | 4000 (2016) | 0 | 4000 (2016) | 0 | 0 | 0 |

TABLE VII. OPTIMAL SOLUTION RESULTS FOR MODEL A AND B (PART B)

| | Cost ($) | | | |
|---|---|---|---|---|
| | PM | corrective | Production | Total |
| Model A | 5645 | 4500 | 48230 | 58375 |
| Model B | 8000 (4032) | 6000 | 49630 | 63630 (59662) |

All parameters are mutual for models A and B, except information about cost and time of PM that is shown in Table 5. We made models A and B to assess and compare the dependency of cost and time of PM to the machine age. In model B, we assume that cost and time of PM are constant and determined with averaging the same parameters in dependency model A. On the other hand, means of cost and time of PM over machine age (index l) are equal for models A and B (see the last column of Table 5). In a practical problem, the cost and time of PM are dependent on machine age. After solving models, A and B, the following result was obtained (Table 6, Table 7). Using mean value and ignoring the dependency assumption in model B causes less planned PM and it causes the corrective cost to go up. Also, the production cost is more than what obtained from model A with a dependency assumption. Considering an independence assumption in model B makes it more unrealistic. To clarify this issue, the following definition is necessary.

G is given as a model objective function for solution X and parameter P. $g^*$ is a value of the objective function for the optimal solution $X^*$ and parameter P, then:

$$g_A^* = G(X_A^*, P_A) \qquad (32)$$

$$g_B^* = G(X_B^*, P_B) \qquad (33)$$

$$\hat{g}_B = G(X_B^*, P_A) \qquad (34)$$

According to Table 6, Table 7, $g_A^* = 58375$ and $g_B^* = 6363$. If we calculate the objective function of model B by using the dependency parameter of model A, which is more realistic, $\hat{g}_B = G(X_B^*, P_A)$ and is equal to 59662 (values in parentheses of Table 6, Table 7). The Model B calculates $g_B^* = 63630$ for a decision-maker at the beginning of a planning horizon, whereas this calculation is unrealistic. True value is obtained by considering the dependency of the time and cost of preventative maintenance as $\hat{g}_B$. In this condition, all costs except the cost of preventative maintenance remain unchanged. In the provided numerical example, this amount has changed from 8000 to 4032. Although, the actual measured $\hat{g}_B$ value is less than $g_B^*$, it is more than $g_A^*$ ($g_A^* < \hat{g}_B < g_B^*$).

Comparing dependency and independence through models A and B shows that assuming the dependency would decrease the total cost. The question is, "How much does dependency effect on decreasing the total cost?" To answer this question, we defined three levels of High, Medium, and Low dependency severity. Each level will be compared to a corresponding case without dependency assumption like what we have done to compare models A and B. Cost and time of PM are presented in Table 8. Differences in each level can be seen in Figure 1.

According to the results shown in Table 9, by increasing the level of dependency, improvement rises. Improvement is defined as $g_A^*/\hat{g}_B \times 100$. The reduction of the total cost for a High level is 8% and it drops to 0.5% for Low-level dependency. Increasing the severity of dependency has led to a greater improvement in the achieved total cost.

We amend periodic constraints to consider periodic PM assumption in model A. Adding periodic constraint does not improve the total cost as shown in Table 10.

Totally, maintenance cost in the noncyclical model is more than a periodic model, but much more reduction in production cost causes the total cost of the cyclical model to get lower than the periodic model.

VIII. CONCLUSION

In this paper, we have proposed a mixed integer linear programming model of integrated multi-product process and maintenance planning (IPPMP) on a capacitated machine that susceptible to random breakdown which leads to noncyclical preventive maintenance and a production model that consider cost and time of PM based on the machine age to make a more realistic model.

TABLE VIII. COST AND TIME OF PM FOR HIGH, MEDIUM AND LOW LEVEL OF DEPENDENCY

| | $l$ | 1 | 2 | 3 | 4 | 5 | 6 | 7 | 8 | Avg. |
|---|---|---|---|---|---|---|---|---|---|---|
| $PMC^l$ ($) | High | 234 | 422 | 760 | 1367 | 2461 | 4430 | 7974 | 14352 | 4000 |
| | Medium | 650 | 974 | 1462 | 2193 | 3289 | 4933 | 7400 | 11100 | 4000 |
| | Low | 1940 | 2327 | 2793 | 3351 | 4022 | 4826 | 5791 | 6950 | 4000 |
| $PMT^l$ (h) | High | 0.2 | 0.4 | 0.8 | 1.4 | 2.5 | 4.4 | 8 | 14.4 | 4 |
| | Medium | 0.6 | 1.0 | 1.5 | 2.2 | 3.3 | 4.9 | 7.4 | 11.1 | 4 |
| | Low | 1.9 | 2.3 | 2.8 | 3.3 | 4 | 4.8 | 5.7 | 6.9 | 4 |

TABLE IX. TOTAL COST RESULTS FOR A DIFFERENT LEVEL OF DEPENDENCY

| | Total Cost | | |
|---|---|---|---|
| | Low | Medium | High |
| $g_A^*$ | 60004 | 54524 | 52042 |
| $g_B^*$ | 63630 | 63630 | 63630 |
| $g_B$ | 60284 | 57578 | 56574 |
| $\frac{g_A^*}{\hat{g}_B} \times 100$ | 99.5% | 94.7% | 92.0% |

TABLE X. COST OBTAINED FOR CYCLICAL AND PERIODIC MODEL A

| | Cost | | | |
|---|---|---|---|---|
| | PM | Corrective | Production | Total |
| Cyclical | 5645 | 4500 | 48230 | 58375 |
| Periodic | 6048 | 4000 | 49630 | 59678 |

The objective function is constituted by two section. The first section of the objective function that should be minimized is related to manufacturing cost including the production cost, holding cost, backorder cost, and setup cost. The second section relates to the maintenance cost including the PM cost and expected corrective cost. It is tried to make a linear model even with binary and integer variables that results the MILP model. The dependency assumption was compared to independency and the obtained results show that considering dependency improves the total cost. Also, by increasing severity of dependency, the total cost reduces more. The presented model can be periodic by amending periodic constraints. In summary, the mathematical model can cope with cyclical and periodic preventive maintenance and dependency and independency of cost and time of PM. The model is developed for a capacitated machine and the model is sought to be expanded for a multistage manufacturing system and present heuristic methods to solve the large-scale problems approximately.

Finally, authors would like to have some recommendation for the interested scholars to expand this research. To proof the practicality of the model, a specific type of manufacturing and machine could be chosen to apply the model e.g., steel convertor plant [5], industrial evaporation network [32], and automotive industries [33]. Furthermore, Including different maintenance strategies adds valuable intuitions to the mathematical model [2], and [34].

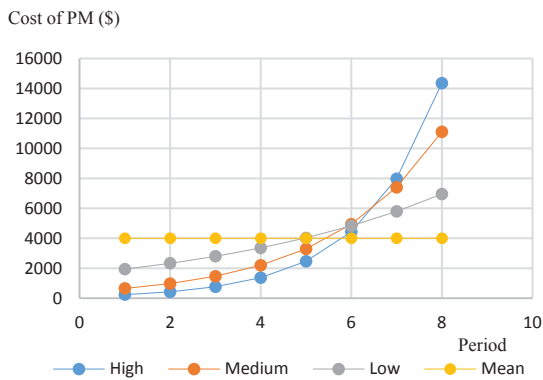

Cost of PM ($)

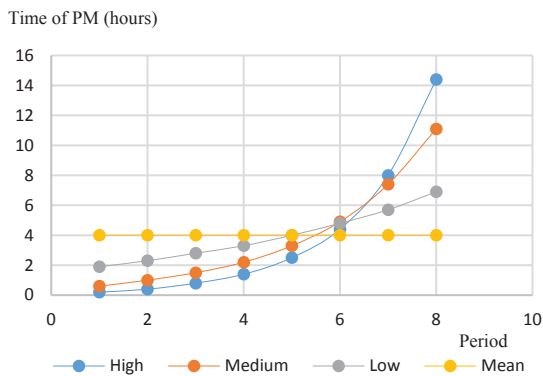

Time of PM (hours)

Fig. 1. Above: Variety of PM Cost based on machine age. Below: Variety of PM Time based on machine age)